\documentclass[prc,aps,floatfix,nofootinbib,twocolumn,letterpaper]{revtex4} 
\usepackage{graphicx} 
\usepackage{amssymb} 
\usepackage{amsmath} 
\usepackage{amsfonts}

\def\be{\begin{equation}} 
\def\ee{\end{equation}}

\begin{document} 
\title{Composite-particle decay widths by the generator coordinate method
}
\author{G.F.~Bertsch}
\affiliation{
Department of Physics and Institute of Nuclear Theory, 
Box 351560\\ University of Washington, Seattle, Washington 98
915, USA} 
\author{W.Younes}
\affiliation{ 
Lawrence Livermore National Laboratory, Livermore, CA 94551, USA }

\begin{abstract}
We study the feasibility of applying the Generator Coordinate Method
(GCM) of self-consistent mean-field theory 
to calculate decay widths of composite particles to 
composite-particle final states. The main question
is how well the GCM can approximate continuum wave functions in
the decay channels.  The analysis
is straightforward under the assumption that the GCM wave functions
are separable into  internal and Gaussian center-of-mass wave
functions.  Two methods are examined for calculating decays widths.
In one method, the density of final states is computed entirely
in the GCM framework.  In the other method, it is determined
by matching the GCM wave function to an asymptotic 
scattering wave function.  Both methods are applied to a numerical
example and are found to agree within their determined uncertainties.
\end{abstract}

\maketitle 

\section{Introduction}

In this work we  propose a simplified computational scheme to calculate
decays of clusters of particles by emission of smaller clusters.
The basic reaction theory has been developed in nuclear
physics following several approaches, most prominently 
the Resonating Group Method (RGM)\cite{na16,su10} and the Generator Coordinate
Method (GCM)\cite{hi53,be03,tr07}.  In the RGM the wave function is expressed as an
antisymmetrized product of internal wave functions of the daughter
clusters together with the relative coordinate wave function between
them.  If there are only a few
particles in each cluster, the antisymmetrization may be
carried out by the use of Jacobi coordinates.
However, that method
scales poorly with the number of constituent particles and is
not practical for large systems.

The GCM is based on a self-consistent mean-field approximation to the many-particle
wave function.  An advantage of 
this approach is that antisymmetrization is
automatic when the system wave function is a Slater determinant of orthogonal
orbitals.  Mean-field theory 
has been quite successful in nuclear physics to describe
binding energies and simple spectral properties of heavy nuclei
\cite{be03}.  
The GCM extends the range of mean-field theory by generating
multiple configurations that can interact with each other as in other 
configuration-interaction methods.  The GCM  introduces
external potential fields into the 
Hamiltonian to  construct the configurations.  
For example, to treat the collective excitations of a cluster, a
single-particle operator would be introduced as a constraining field.
The wave function basis would include some configurations
for which the expectation values of the operator would sample
the range of variation in the physical excitation.

The application of the GCM to reactions involving clusters also has
a long history in nuclear
physics\cite{ku69,ho70,de72,ha72,ta72,ho73,fr75,be75,hu77,on79},
but with less success up to now.  One problem was the large size
of the single-particle space needed to 
adequately represent a configuration of separated daughter clusters.  
Fortunately this is no longer an issue with present-day computer resources
\footnote{ See for example Ref. \cite{bu16} for present-day
capabilities.}.
More fundamentally, a problem that still has no clear solution
is how to treat the relative coordinate between daughter clusters
in the decay channel.  Asymptotically the wave function must
factor into a product of the internal wave functions of the clusters
and a one-dimensional wave function of the relative coordinate as
in the RGM.
However, in mean
field theory the center of mass is just a wave packet and
not a true coordinate.  How to join the two representations (RGM and GCM)
has been the subject of much of the literature.
   
Our goal in the present work is not so ambitious as to develop
a full reaction theory for large clusters.  
Rather, we focus on the more modest problem of calculating rates of
decay into cluster channels.
In fact decay rates were hardly discussed 
in the early theory, apart from semiclassical treatments of alpha-particle
decay.
  
Our approach is through Fermi's Golden Rule formula,
\be
\label{FGR}
\Gamma(i \rightarrow f) = 2 \pi   \langle i | H | f\rangle^2
\frac{ d n_f}{ d E}.
\ee 
Here $i$ is the initial mean-field configuration.  For example,
we have in mind a self-bound excited state of the 
parent cluster. The final state $f$ is the unperturbed wave function in
the decay channel at the same energy.  It will be
mostly represented on a finite basis of GCM configurations
in which the relative coordinate has been constrained
to a mesh of discrete values.
The last factor is the density of final states in the 
$f$ channel. One method to determine it is
to join the GCM wave function to the RGM 
scattering wave function. Typically the wave functions are
matched at a point
$R$ selected to be somewhat outside the distance where the clusters
touch.  

\begin{figure}[tb] 
\begin{center} 
\includegraphics[trim = 5cm 5cm 0 1cm, clip=True,width=1.6\columnwidth]{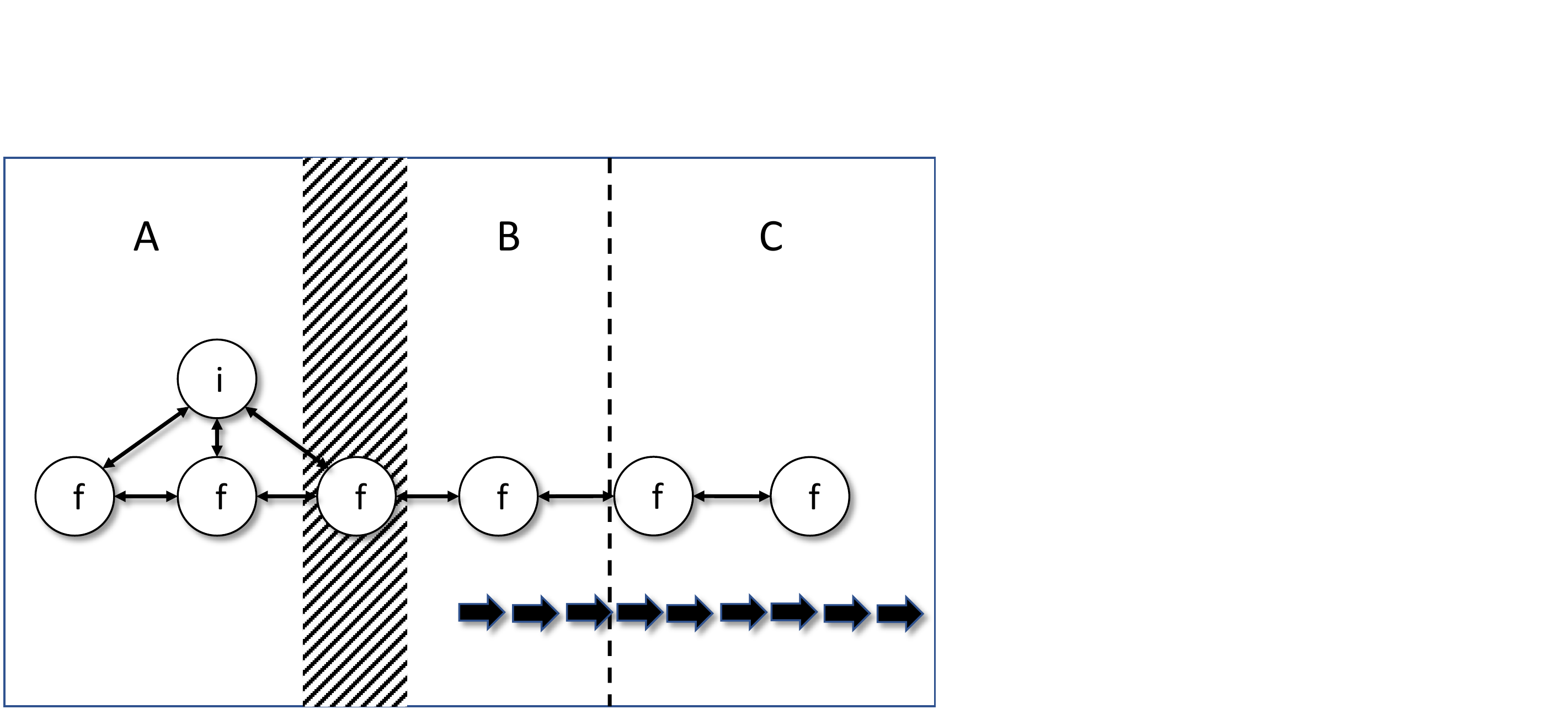}
\caption{Schematic view of the wave functions involved in
calculating decay widths by the GCM. See text for
explanation.
\label{diagram}    
}
\end{center} 
\end{figure}

Pictorially, the relationship between 
configurations and wave functions 
is shown in Fig. \ref{diagram}.
The horizontal axis is a generator coordinate 
that includes
fused or strongly interacting configurations (region A) as
well as regions of separated clusters (regions B and C).  In regions
B and C the coordinate could be the operator measuring the separation
of the clusters, Eq. (\ref{zrel}) below.  In region A or B the operator
could be some other measure of shape such as the quadrupole moment
operator. In 
the regions of separated clusters, we require that the asymptotic
RGM wave functions are valid without any need for antisymmetrization
between cluster.  The line of horizontal arrows indicates that
asymptotic relative coordinate wave function.
The vertical line between
region B and C is the chosen matching point between the
two representations.  Finally, the configuration $i$ in region
A is the initial state whose decay width is the object of the
theory.  

In Sect. II below, we explore from a computational
point of view the fidelity with which
the relative-coordinate wave in the asymptotic region
can be represented in a discrete
basis of GCM configurations.  Characteristics that can be
compared are wave function overlaps, eigenstate energies, and
logarithmic wave function derivatives.
Sect. III deals with calculating $d n_f/ dE$.
It will be seen that a simple approach without the
RGM wave functions is sufficient for rough estimates.  However,
when there are strong long-range potential fields in the final
state, matching to the asymptotic RGM is unavoidable.  We assess the accuracy
of that procedure by determining its  sensitivity to the choice of
matching point $R$ and to the parameters defining the GCM 
configuration space.

\section{Continuous wave functions from a 
discrete basis}

We are interested in the accuracy of relative-coordinate
wave functions obtained from a discrete GCM basis.
The problem of
representing the center-of-mass wave function in a discrete basis of
single-cluster GCM is nearly identical, and in this Section
we simplify the notation 
accordingly.
We start with a translationally invariant Hamiltonian $H$ that
can be solved in the mean-field approximation to produce
many-particle configurations $\Psi_{gcm}^{\alpha}$.
These wave functions have the form of Slater
determinants.  External one-body fields $Q$ have been 
added to the Hamiltonian, with the strength of the fields
adjusted to produce desired expectation values 
$\langle Q \rangle$, and the label $\alpha$ in 
$\Psi^\alpha_{gcm}$ includes this information.
For the cm position of a single cluster containing $N$ particles, the field 
would obviously be $\vec r/N$.  
For the relative
motion of two clusters along the $z$-axis, one can choose a 
dividing plane perpendicular to the axis located at some point
$R$.
The constraining operator is 
\be
\label{zrel}
z_{rel} = (z-R)\Theta(z-R)/N_R + (R-z)\Theta(R-z)/N_L
\ee
where $N_R,N_L$ are the number of particles on each side.

\begin{figure}[tb] 
\begin{center} 
\includegraphics[width=1.0\columnwidth]{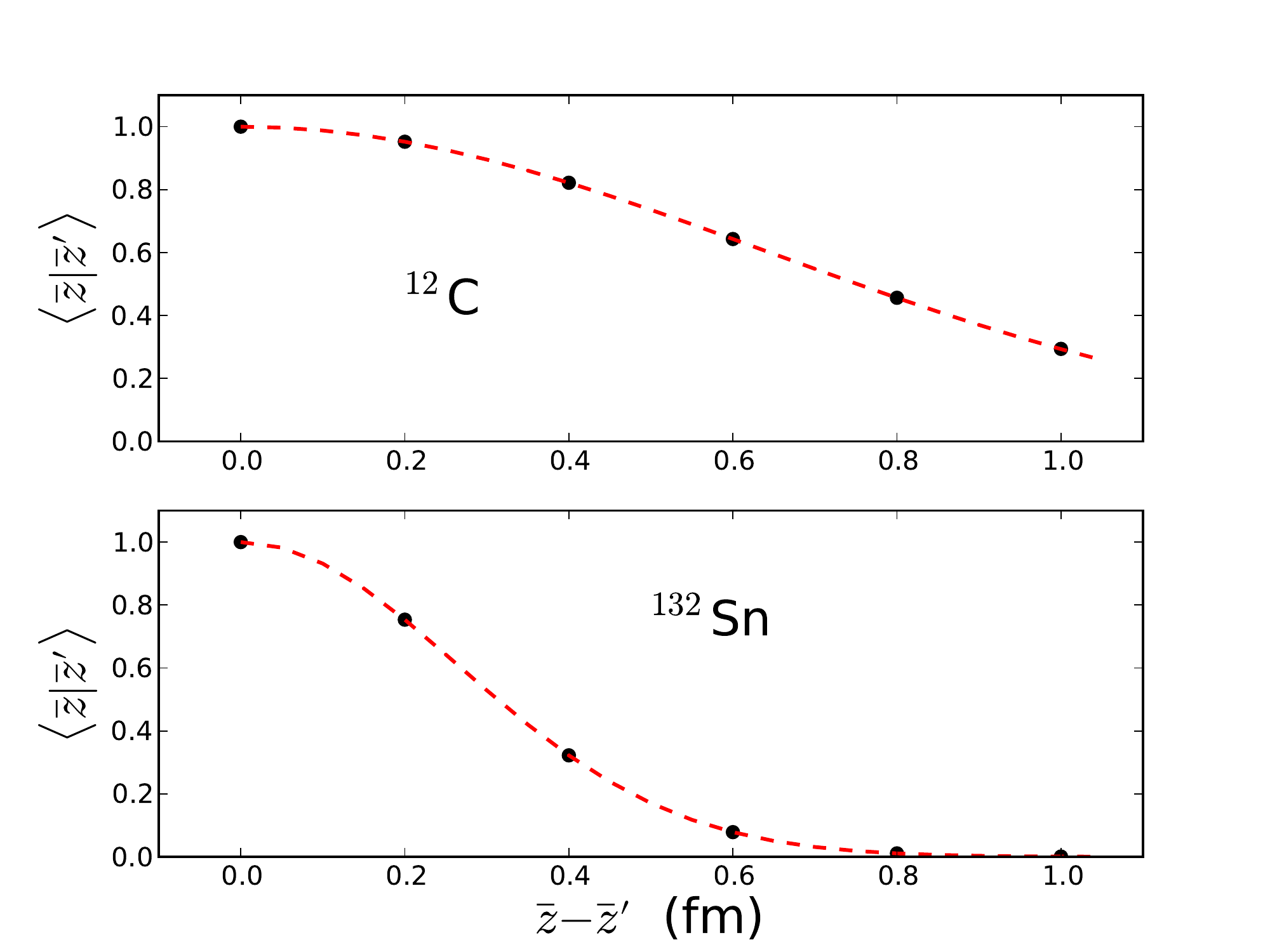}
\caption{Overlaps of the nuclear $^{12}$C and $^{132}$Sn mean-field  
wave functions as a function of cm coordinate displacement.  
Circles: overlaps of the GCM configurations; Dashed line:
fit to a Gaussian function.
Wave functions were calculated in a harmonic oscillator space containing
12 complete shells, using the Gogny D1S energy functional with no
cm energy correction.
\label{gauss-fit}    
}
\end{center} 
\end{figure}

We assume that 
the GCM wave function of a single
cluster
$\Psi^\alpha_{gcm}$ can be factorized into an internal wave function
$\Psi^\alpha_{int}$ times a center-of-mass wave function $\psi_{cm}$,
\begin{equation}
\label{separable}
\begin{aligned}\Psi^\alpha_{gcm}\left(\vec r_{1},\vec r_{2},\ldots\right) &
=\Psi^\alpha_{int}
\left(\xi\right)\psi^\alpha_{cm}\left(\bar z_\alpha, z_{cm}\right)\end{aligned}
\end{equation}
Here $\vec r_{1},\vec r_{2},\ldots$ are the coordinates of the constituent
particles, $z_{cm}$ is a center-of-mass coordinate, and $\xi$ are unspecified
internal coordinates.
The parameter $\bar z_\alpha$ is the expectation value  
$\bar z_\alpha = \langle \alpha | z_{cm}| \alpha\rangle$ 
Factorization is a strong assumption, but there is some justification for it
in nuclear theory.
As was noted in some of the cited references, Eq. (\ref{separable}) 
is exact for the ground state
of a many-particle system
in a harmonic oscillator potential.
Indeed, in the early studies
the wave function were assumed to be harmonic oscillator eigenstates
and thus factorizable.
In a more general GCM treatment, information about the center-of-mass 
coordinate 
can be obtained by taking the overlap of the wave functions  under
displacement.  It is an empirical fact that the overlap functions
are close to Gaussian,
\be
\int d z_{cm} \, \psi^{\alpha*}_{cm}\left(\bar z_1,z_{cm})\right)
\psi^\alpha_{cm}\left(\bar z_2,z_{cm}\right)  
\ee
\be
\approx \exp(- (\bar z_1-\bar z_2)^2/4s^2)
\ee
for some size parameter $s$.
Two examples from nuclear physics are shown in Fig. \ref{gauss-fit}.
The nuclei differ in particle number $N$ by an 
order of magnitude, but the length parameter $s$ in the fitted 
Gaussians differ only by a factor of $2.5$.  
Combining the factorization assumption together with the observed
near-Gaussian overlaps,  the normalized cm wave function in Eq.
(\ref{separable}) is given by
\be
\label{gauss}
\psi_{cm}(\bar z,z_{cm})
 = \left(\frac{1}{ \pi s^2}\right)^{1/4}
e^{-(\bar z-z_{cm})^2/2 s^2}
\ee

The wave functions of physical interest are the stationary states in the space of the
GCM configurations.  These have the form
\be
\begin{aligned}\Psi^\lambda_{gcm} & =
\sum_{\alpha}a_{\alpha\lambda}\Psi_{gcm}^{\alpha}\end{aligned}
\label{eq:Psi}.
\ee
where $a_{\alpha\lambda}$ is an amplitude
and $\lambda$ is a label to distinguish the eigenstates.
The amplitudes are obtained from the solutions of the
non-Hermitian eigenvalue
problem \cite{wi65}
\be
\sum_{\alpha}H^{gcm}_{\alpha'\alpha}a_{\alpha\lambda} 
=E_{\lambda}\sum_{\alpha} N^{gcm}_{\alpha',\alpha} a_{\alpha\lambda}.  
\label{eq:Egcmf}
\ee
Here $H^{gcm}$ and $N^{gcm}$ are the Hamiltonian and overlap matrices in the
GCM basis. The amplitudes $a$ are normalized as
\be
\sum_{\alpha,\alpha'} a_{\alpha,\lambda}N^{gcm}_{\alpha,\alpha'}
a_{\alpha',\lambda'}  = \delta_{\lambda,\lambda'}
\ee
The machinery to calculate Eq. (\ref{eq:Egcmf})  is
well developed\cite{be03} and will not be discussed here.  Suppose that the GCM basis
states are all in the asymptotic region and the configurations are
constructed on a uniform mesh $\bar z_\alpha = n\Delta z$.
Then we can drop the subscript $\alpha$ on $\psi_\alpha$ and write
\be
\label{psiz-gcm}
\psi^\lambda(z) = \sum_{n=n_i}^{n_f}  a_{n,\lambda} \psi(n \Delta z,z)
\ee
where $n$ is an integer in the range $n_i,n_f$.  We now 
examine how well this wave function (and the associated eigenenergy
$E_\lambda$) reproduces the exact $\psi(z)$ obtained by solving the
Schr\"odinger equation for the RGM center-of-mass coordinate.  

The most important parameter in the method is the mesh spacing;
the accuracy that can be achieved with Eq. (\ref{psiz-gcm}) depends
on the dimensionless ratio   $\Delta z / s$. 
There are two conflicting demands in the choice of mesh parameter. If 
$ \Delta z >> s$, the spacing will be too sparse to approximate
the continuum wave functions.  On the other hand, if
$ \Delta z << s$ the GCM space will be effectively overcomplete
and the norm matrix $N^{gcm}$ will be nearly singular.  The choice
\be
\label{dz-def}
\Delta z = 5^{1/2}s
\ee
appears to be a reasonable compromise and we use is for most of the numerical
examples.  But one of the methods we examined to calculate decay width
requires a somewhat finer mesh, as will be seen in Sec. III.

\subsection{Plane waves}
\begin{figure}[tb]
\begin{center} 
\includegraphics[width=\columnwidth]{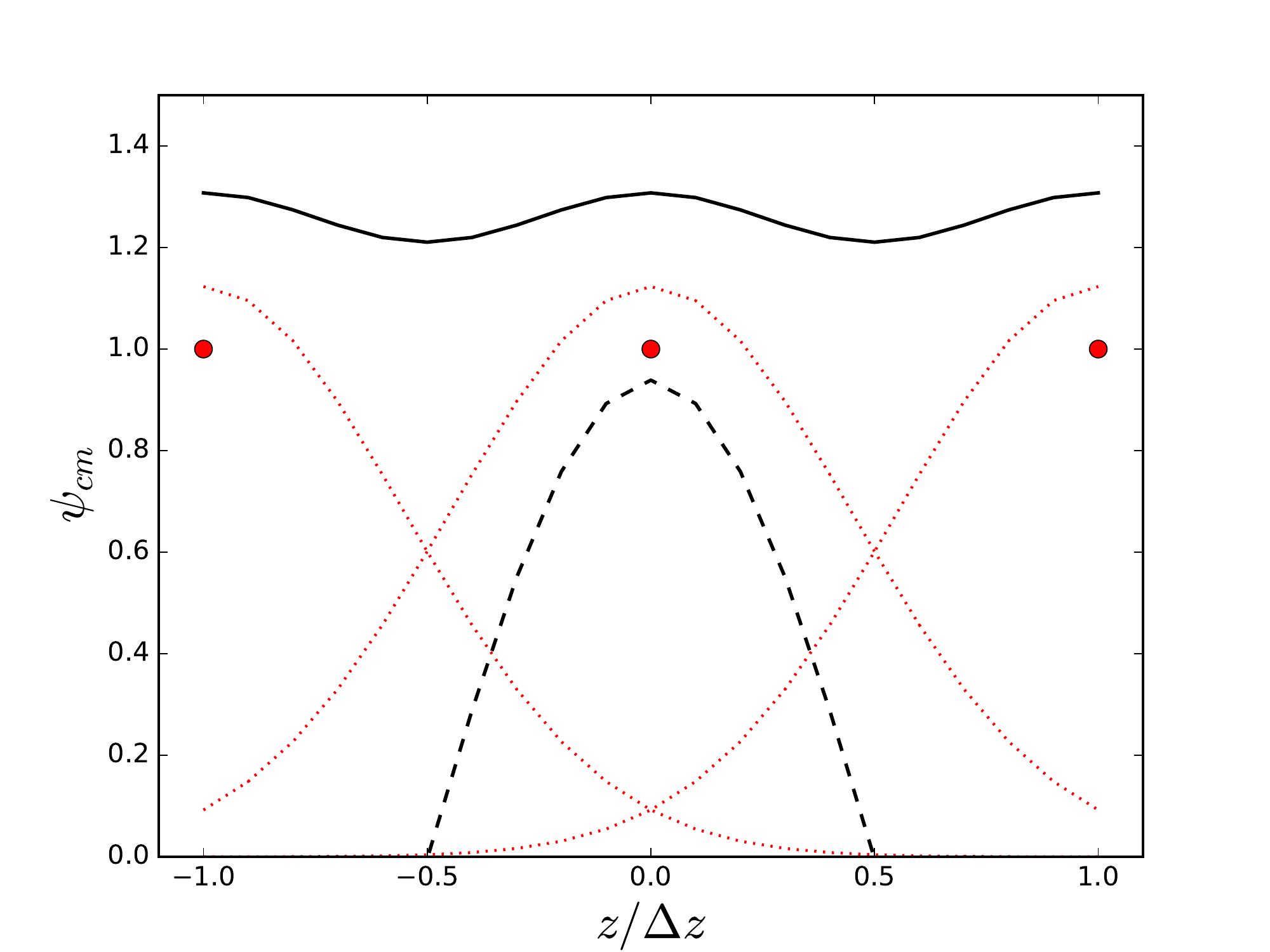}
\caption{CM wave functions for a uniformly spaced basis in the CM
coordinate, in units of $\Delta z$.  The red circles are the amplitudes
$a$ at the mesh points.
Dotted lines are from individual GCM configurations; solid line is the 
approximate $k=0$ wave function; dashed line shows the positive part
of the wave function for $k=\pi/\Delta z$.
}
\label{wf_cm}    
\end{center} 
\end{figure}

We start with a free-particle Hamiltonian on the infinite 
interval $z=(-\infty,\infty)$ and a GCM basis defined by Eq.
(10-\ref{dz-def}).
with the mesh space Eq. (\ref{dz-def}).  
By translational symmetry the GCM eigenstates can be expressed as
\be
a_{n,k} = e^{i k \Delta z}
\ee
where $k$ is in the interval $(-\pi/\Delta z, \pi/\Delta z)$. 
The resulting  wave function is 
\be
\label{ft}
\psi_k(z) = \sum_n e^{i k \Delta z} \psi_{cm}(n \Delta z, z).
\ee
It should represent a plane wave of momentum $k$.  As
an example, Fig. \ref{wf_cm} shows the components $\psi_{cm}(n \Delta z, z)$
in the range $(-\Delta z,\Delta z)$ and the wave function $\psi_k(z)$
for $k=0$ and $k=\pi/\Delta z$.
Visually, the $k=0$ function (solid line) is quite flat, showing that
it is close to a zero-momentum eigenstate.  Of course there is a 
residual variation of the wave function due to the discrete
basis.  In the range of discretizations considered here, the
relative variation can be estimated from the Poisson summation formula
as
\be
\frac{\psi}{\overline \psi} \sim 1 \pm  2 e^{-2 \pi^2 s^2/\Delta x^2}
\ee
where $\bar \psi$ is the average value of $\psi(z)$.

The figure also shows (dashed line) the positive part of the wave function for the
maximum momentum contained in the basis, 
$k=\pi/\Delta x$.  It is close to 
cosine function of argument
$\pi z / \Delta z$, apart from normalization.  Note that the corresponding sine
function cannot be represented in the basis.

\begin{figure}[tb]
\begin{center} 
\includegraphics[width=\columnwidth]{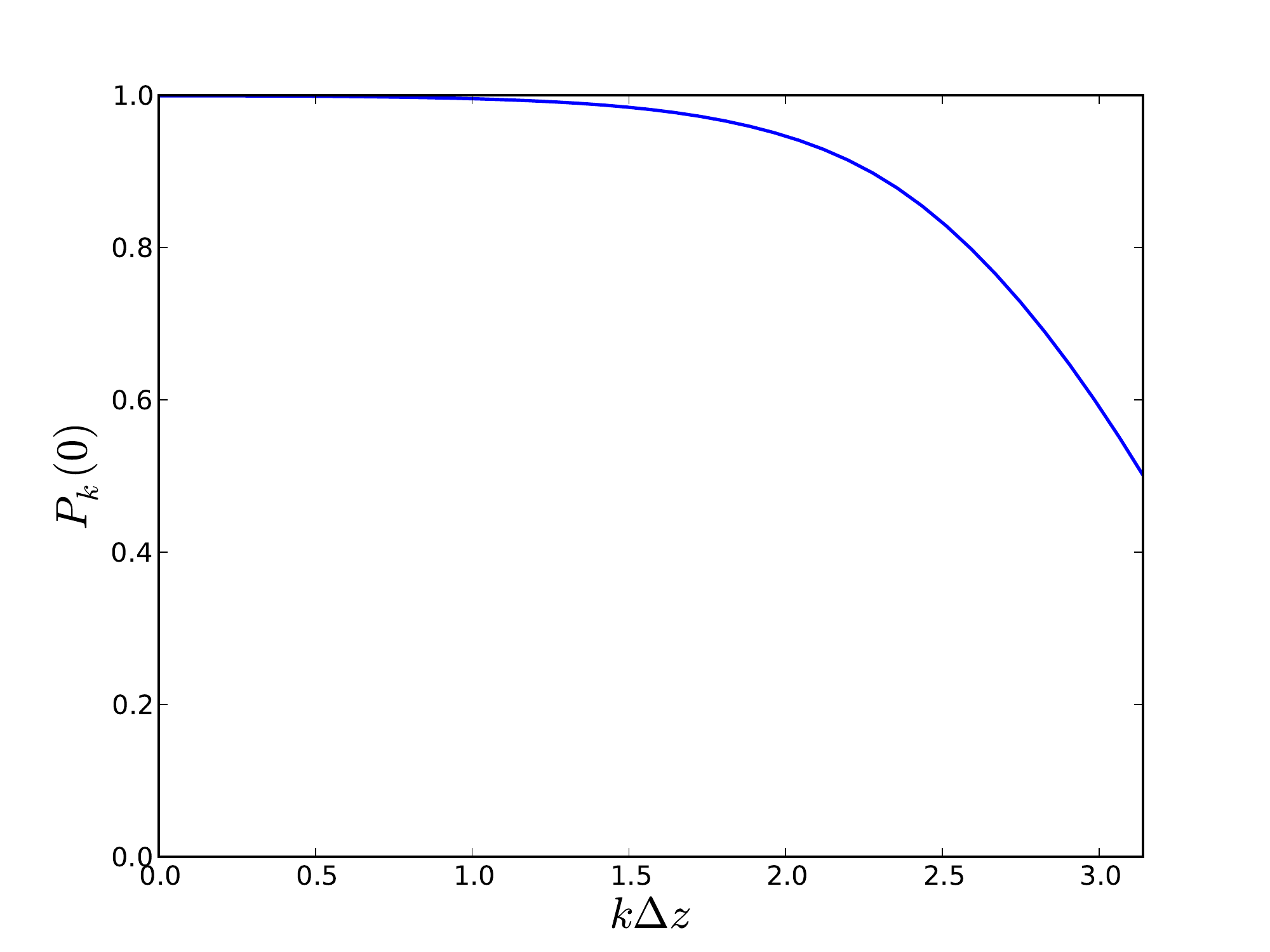}
\caption{
\label{Pkm}    
Probability of momentum $k$ in the GCM approximation $\psi_k$.
}
\end{center} 
\end{figure}

For a quantitative measure of the fidelity of the GCM representation,
one can calculate the overlaps with true
momentum eigenstates by Fourier transform.  
The probability $P_k(m)$ of 
momentum $k_m = k + m 2 \pi/\Delta z$ can be computed as 
\be  
\label{Pm}
P_k(m) = \frac{\left|\int_0^{2 d z \,\Delta z} e^{-i k_m z} \psi_k(z)\right|^2}{
\int_0^{2 \Delta z} dz \,\left|\psi_k(z)\right|^2}
\ee 
Fig. \ref{Pkm} shows $P_k(0)$ over the range $k=(0,\pi/\Delta
z)$. One sees that $P_k(0)$ is close to one up to $k \approx \pi/2\Delta z$.  Beyond
that the representation becomes poorer; 
at the
upper limit it approaches $1/2$, with 
the $m=-1$ Fourier component taking nearly all of the remaining strength.  

Another test of the representation is how well it reproduces
the plane-wave energy spectrum,
\be
\label{energy}
E_k = \frac{\hbar^2 k^2}{ 2 M}.
\ee
Here $M$ is the mass of the cluster.  
The energy can be calculated the
ratio of expectation values  
\be
\label{Egcm}
E^{gcm}_k = \frac{\langle k |H^{gcm}| k \rangle}{
\langle k | N^{gcm} |k \rangle}. 
\ee
The results for the  
numerator and denominator are
\be
\langle \psi_k | N^{gcm}| \psi_k \rangle =  N \left( 1 + 2 \sum_{n=1}^\infty\cos(n k 
\Delta z)
N^{gcm}_{0n} \right) 
\ee
and
\be
\langle \psi_k |H^{gcm} | \psi_k \rangle =  
N \left( H^{gcm}_{0n} + 2 \sum_{n=1}^\infty\cos(n k \Delta z)
H^{gcm}_{0n} \right)
\ee
where $N$ is the number of basis states.
The required overlap matrix 
elements  are given by
\be
N^{gcm}_{0n}  = \exp(- n^2(\Delta z /2s)^2).
\ee
The matrix elements for the kinetic energy operator 
\be 
T = -\frac{\hbar^2}{ 2 M} \frac{\partial^2 }{ \partial z^2_{cm}}
\ee
 are
\be
T^{gcm}_{0n}
  = 
N^{gcm}_{0n}  E_0
\left( 1 - \frac{(n \Delta z)^2
}{ 2 s^2}\right). 
\label{0Kn}
\ee
where 
\be
E_0 = \frac{\hbar^2}{  4 M s^2} 
\ee
is the expectation value of
the kinetic energy in the wave function $\psi_{cm}$.
$E_0$ is an important parameter setting the energy scale for
the validity of the GCM basis as formulated here.

\begin{figure}[tb]
\begin{center} 
\includegraphics[width=\columnwidth]{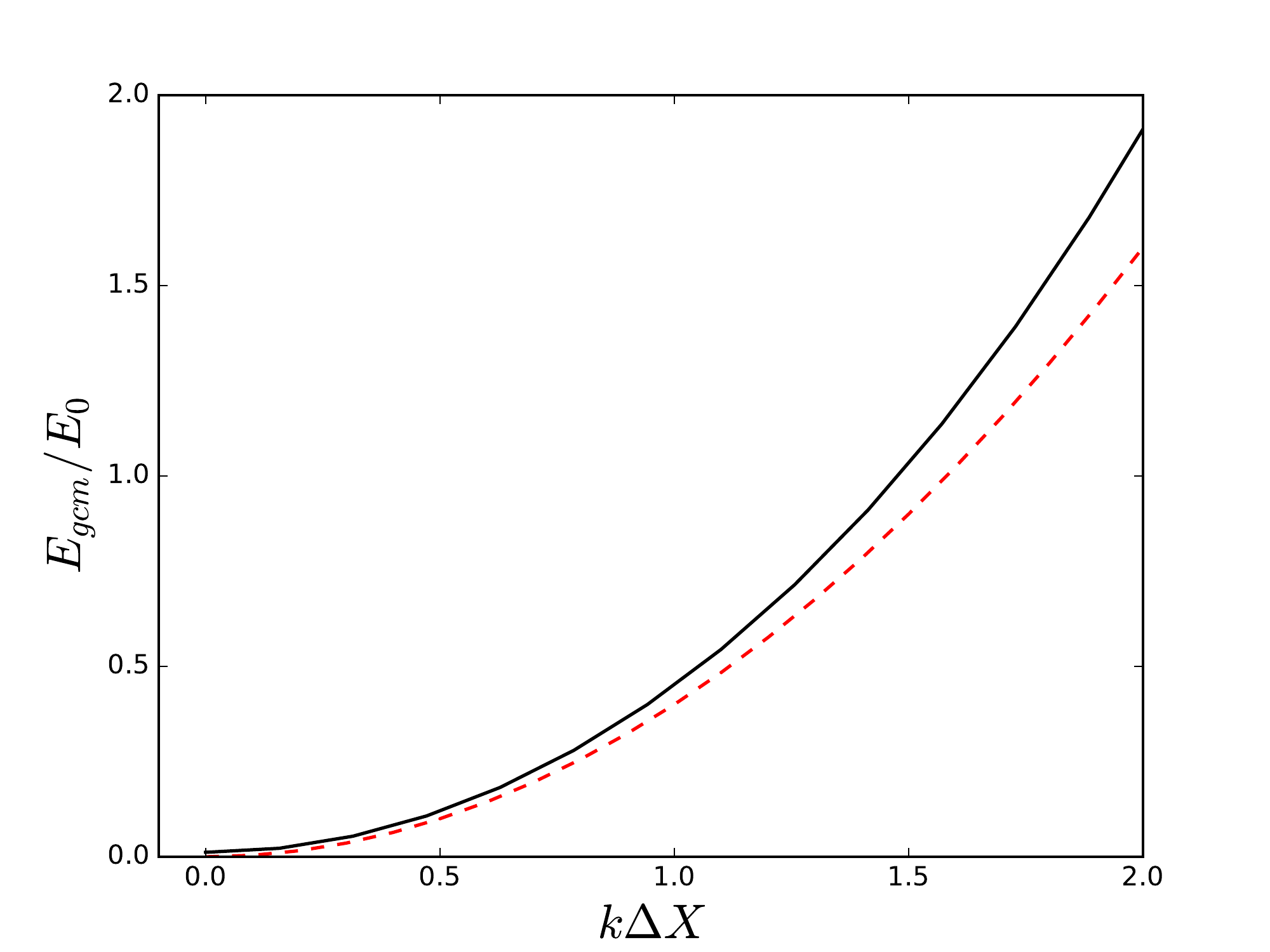}
\caption{Solid line:  free-particle energy versus momentum in the
GCM approximation as discussed in the text; dashed line:  exact
energy from Eq. (\ref{energy}).
}
\label{dbE}    
\end{center} 
\end{figure}

The accuracy of the GCM kinetic energy Eq.~(\ref{Egcm}) under the
conditions of the previous example may be seen in Fig. \ref{dbE}.
The dashed line is the exact energy (Eq. (\ref{energy})) and the solid
line is the GCM result.  There
is a slight offset at $k=0$, but apart from that 
the error is less than 15\% up to 
$k \approx \pi/2\Delta z$.
We judge the fit to be quite good for estimates not requiring
wave function matching.

The wave-function matching can be carried out by renormalizing the
$\psi(z)$ to 
reproduce both the amplitude and logarithm derivative of the
asymptotic scattering wave function at $R$.  Some preliminary indication 
of the error associated with this procedure
be seen in Fig. 3:  $\psi_k$ at $k=0$ undulates 
with an amplitude of about $4$ \%.  This suggests
that normalization obtained by matching at different points
$R$ would vary by a similar amount.  Since the decay rate is
quadratic in the normalization factor, this would cause
an 8 \% uncertainty in the calculated rate.  
This source of error will be
treated in more detail in Sec. \ref{potentials}.
 
\subsection{Potential fields}
\label{potentials}
We now add a potential $V$ to the Hamiltonian,  
with $V$ depending only on 
$z_{cm}$.  The GCM matrix elements are computed as
\be
V^{gcm}_{n,n'} = \int dz \psi^*(n\Delta z, z) V(z) \psi(n'\Delta z,z)
\ee
to give a Hamiltonian $H^{gcm}_{n,n'}= T^{gcm}_{n,n'}
+ V^{gcm}_{n,n'}$.
The energy scale
$\hbar^2/2M \Delta z^2$ will set the permissible range of variation in $V$
when approximating the continuum wave functions.
It is easy to show \cite{RS} that the GCM representation is
exact for a harmonic oscillator potential in the limit $\Delta z \rightarrow
0$.

\begin{figure}[tb]
\begin{center} 
\includegraphics[width=\columnwidth]{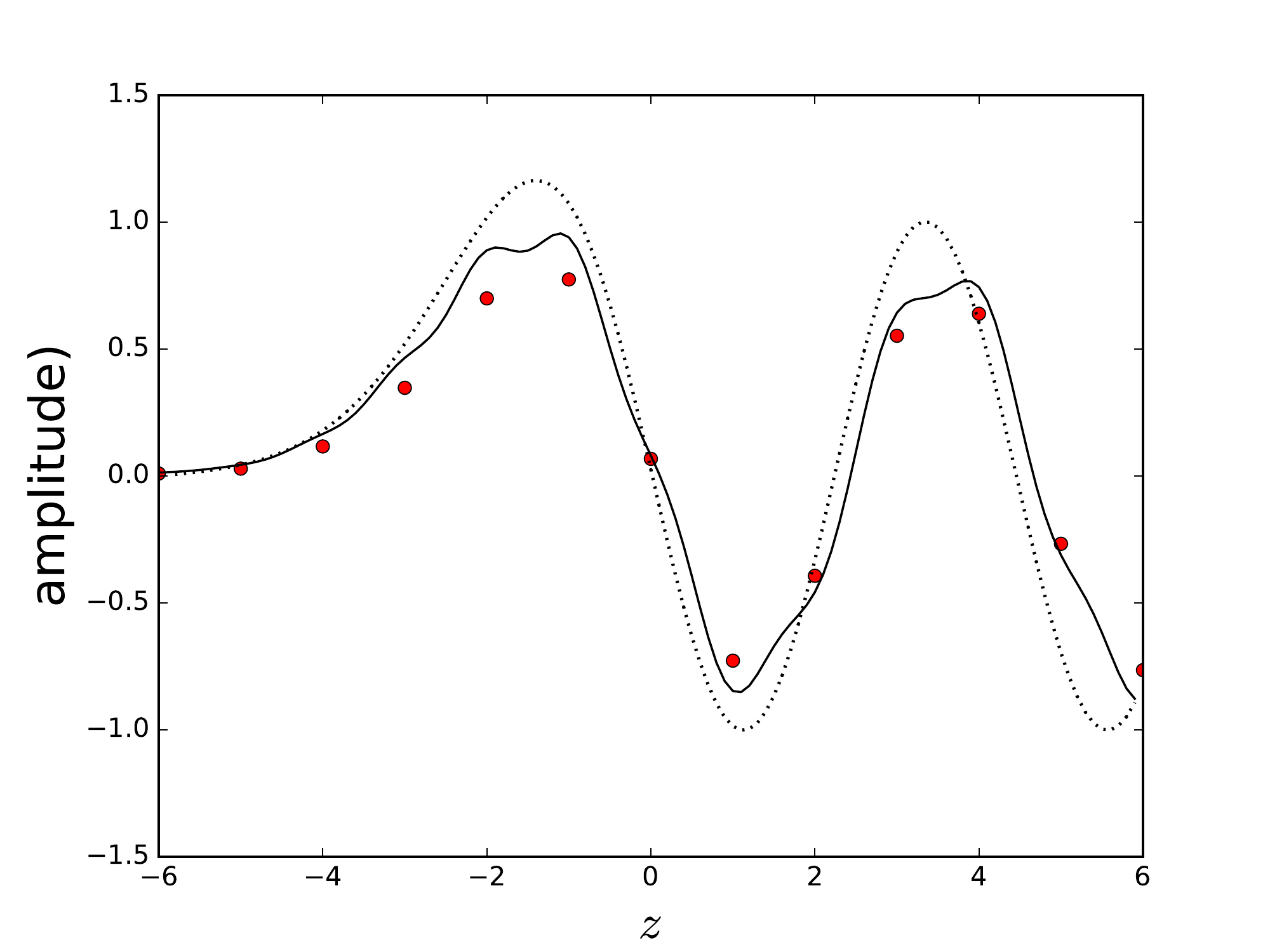}
\caption{
\label{ramp}    
Comparison of the GCM and Schr\"odinger wave functions
for the ramp Hamiltonian Eq.(\ref{rampV} with $F = 0.4$.
The GCM wave function is
constructed in a basis of 13 configurations centered on the mesh
$(-6\Delta z, 6\Delta z)$. The circles show the amplitudes
$a_{n,\lambda}$ for the third excited state at energy $E = 1.029 $.
The corresponding $\psi_\lambda(z)$, shown as
the solid line,
is computed from Eq. (\ref{psiz-gcm}) taking $s=\Delta z/5^{1/2}$.
Amplitudes and wave function have been scaled by a factor of 2
to facilitate comparison with the Schr\"odinger function, shown
as the dotted line.  That wave function is sinusoidal for $z>0$
and has been normalized to  $\sin( k z +\delta)$ for positive $z$.
}
\end{center} 
\end{figure}

We examine here the performance of the GCM taking $V$ to be
a linear ramp potential in the negative $z$ region
\be
\label{rampV}
V(z) = F |z|  \Theta(-z).
\ee 
where $F$ is a positive constant.  The solutions to the Schr\"odinger
equation for $H$ will be sinusoidal for $z >0$ and  
decay as a scaled reflected Airy function
for large negative $z$.  Fig. (\ref{ramp}) compares the
Schr\"odinger and the GCM wave functions for the set of 
parameters given in the caption.    
One sees that the GCM wave function roughly follows the sinusoidal
form of the Schr\"odinger solution, but there are small unwanted
undulations similar to those seen in Fig. 3.  They are an artifact
of the finite mesh spacing and can be reduced by decreasing it.
\begin{figure}[tb]
\begin{center} 
\includegraphics[width=\columnwidth]{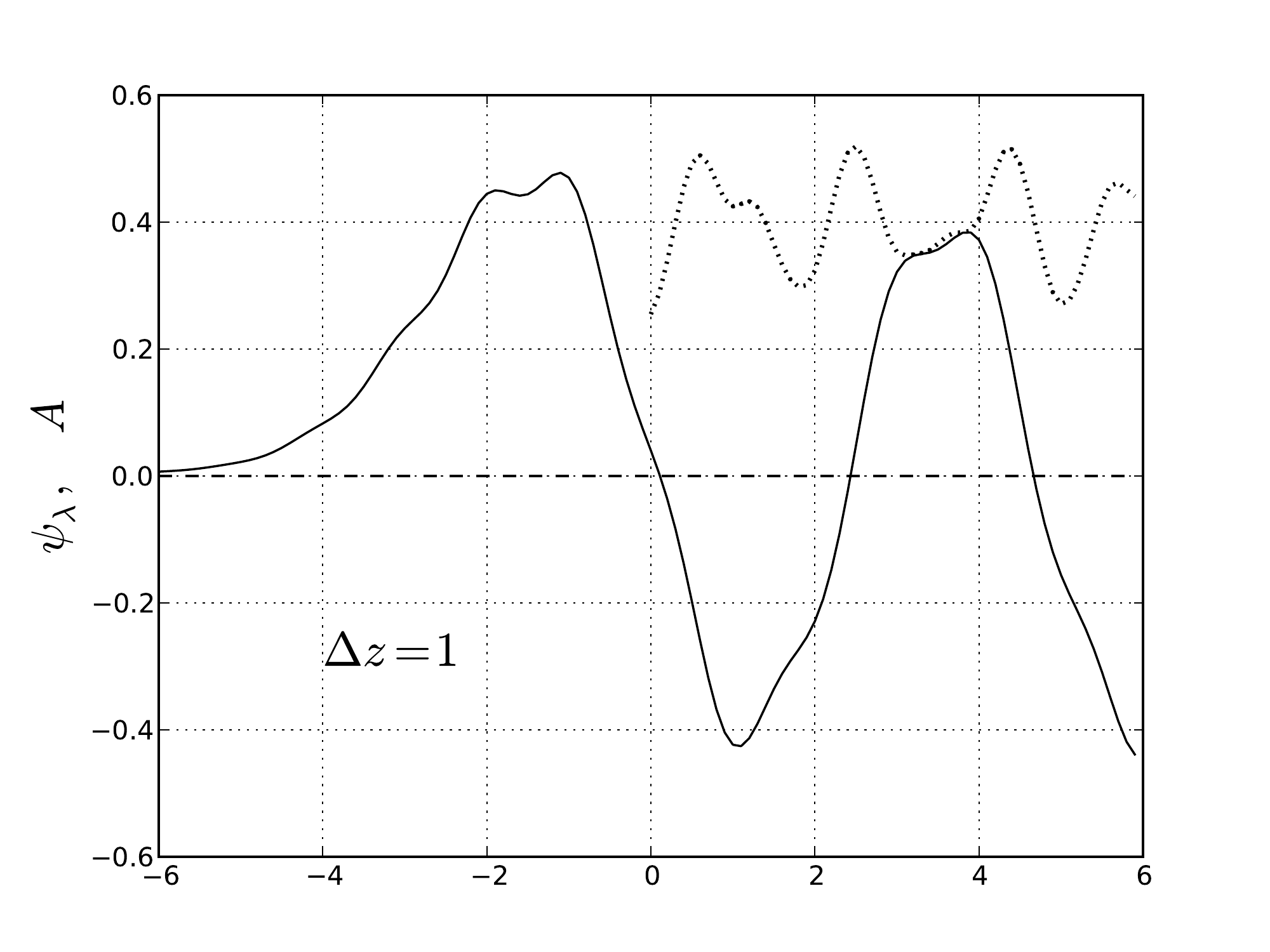}
\includegraphics[width=\columnwidth]{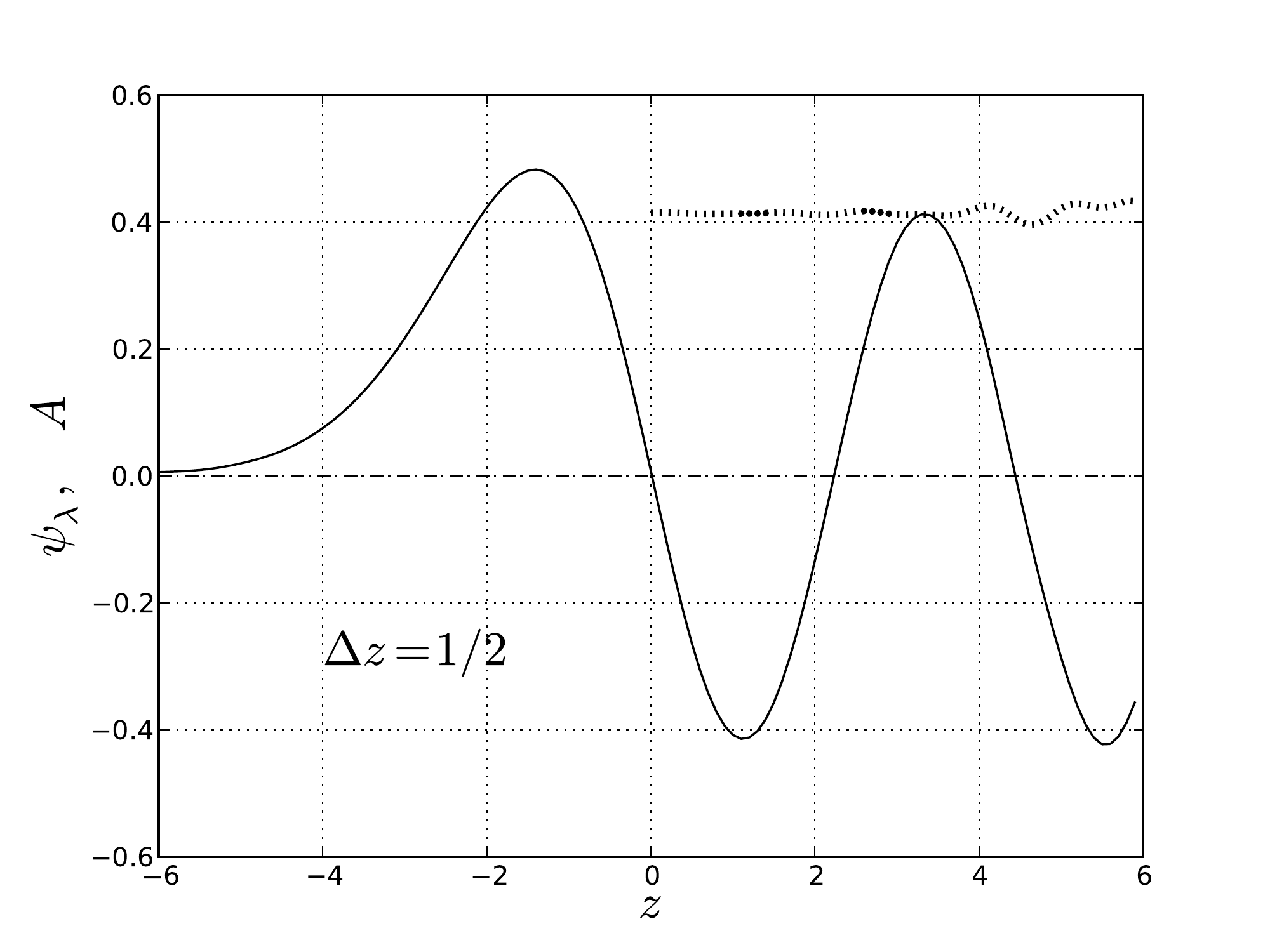}
\caption{
\label{rampA}    
Solid line:  wave function $\psi(z)$ for the ramp Hamiltonian.
Dotted curve:  sine function amplitudes $A$ from Eq. (\ref{A}).  Top and
bottom panels show the results for mesh spacing of $\Delta z = 1$ and
2, respectively.  In the
top panel, the variation in $A$ is $0.30-0.42$ for the
range  $R = 1-3$.  The variation reduced to $ 0.411  < A < 0.418 $ in the
bottom panel.
}
\end{center} 
\end{figure}

The critical test of the numerical approximations is how well
the normalization of the GCM can be determined when matching
to the Schr\"odinger solution.
Assume that the GCM wave function has
the form $ \psi_\lambda(z) = A \sin(k z + \delta)$ at the chosen
matching point $R$.  Then
$A$ is given by
\be
\label{A}
A = \psi_\lambda(R) \frac{(k^2 + {\cal L}^2)^{1/2}}{k}
\ee
where $\cal L$ is the logarithmic derivative
\be
{\cal L} = \frac{1}{\psi_\lambda(R)} \left.\frac{d \psi_\lambda}{d
z}\right|_R.
\ee   
Fig. \ref{rampA} shows the amplitude $A$ as a function of $R$ calculated 
this way. One sees that it  fluctuates over a range of
about 30\% depending on the choice of $R$.  The decay formula requires the 
square of the asymptotic amplitude, so the uncertainty
in the calculated decay width will be as much as a factor of two.  
Clearly one would like to do
better than this.  One way is to decrease the mesh spacing, but
there may be other ways based on properties of the
unwanted undulations.

\section{Formulas for the cluster decay widths}

Under the factorization Ansatz, the
GCM wave function for a configuration of two separated clusters will have
a product of the individual cm wave functions
$\psi_{cm1}(z^1_{cm})\psi_{cm2}(z^2_{cm})$.
Furthermore under the Gaussian assumption, that wave function can be written as 
a product of a Gaussian for the relative coordinate $z_{rel}=
z^1_{cm} - z^2_{cm}$ times a Gaussian for another linear combination
of $z^1_{cm}$ and $z^2_{cm}$.  Thus the relative coordinate can
be separated out and treated in exactly the same way as was done
for $z_{cm}$ in the last section.  Of course the mass $M$ in the
kinetic energy of the final state is now the reduced mass of the
two-cluster system.

We now return to Eq. (\ref{FGR}).  The state $i$ can
be any configuration of the parent cluster that is
stable under the mean-field Hamiltonian, with one qualification
mentioned below.
The $f$ channel is defined in the external region by the
mean-field configurations of individual isolated daughter clusters.
The
channel needs to be defined in the internal region (A) as well.
For this purpose, it would be helpful to introduce additional
constraints to ensure that the added configurations 
are the ones with the largest Hamiltonian matrix elements
connecting to the B-region configurations.
For example, one could demand the basis be constructed using
axially symmetric 
mean-field Hamiltonians.
Then 
the orbitals are characterized by their angular momentum
projections $J_z$ about the $z$-axis.  The Hamiltonian matrix elements will 
be those which do not change the orbital occupancies with respect to 
$J_z$.  A specific example is given in the Appendix; see also
Ref. \cite{be18}.
We note that an axial basis has been employed in
chemical reaction theory to simplify the treatment of the
interaction \cite{re74}. Also, the conservation of orbital
symmetry is an important principle for understanding organic
reactions \cite{wo65}. 

Let us assume now that the GCM basis has been constructed
for the $f$-channel chain and $H^{gcm}$ has been diagonalized to
obtain eigenstates and their energies.   The spectrum 
will be discrete since the basis is finite.  This raises
a technical issue  in that the $f$ eigenstate should have the 
same energy as the initial state $i$.
It would be straightforward to add a diagonal
term to $H^{gcm}$ to tune the energy of one of the eigenstates
to match $E_i$.  If only matrix elements exterior
to the matching point $R$ are adjusted, it shouldn't matter
how it is done.  
One last point is that the states $i$ and $f$ should be
rigorously orthogonal; otherwise the perturbation formula
Eq. (\ref{FGR}) cannot be directly applied.
Note that orthogonality
is automatic the occupation numbers are
different in a basis having an orbital symmetry.

In the
numerical example below, we will also assume that the 
center and spread of the relative coordinate wave function of $i$
is the same as that of one of the $f$-channel configurations, 
say $n=n_c$. Then the
matrix
elements between $i$ and the $f$-channel configurations
can be expressed as
\be
H^{gcm}_{i,n} = v_0 N^{gcm}_{n_c,n}
\ee
where $v_0 = H^{gcm}_{i,n_c}$.
It should be emphasized that this assumption is only made
for numerical convenience here; in practice the $(i,n)$ matrix elements
would be calculated in the usual way using the GCM machinery.
The expression for the squared interaction matrix element in Eq. (\ref{FGR})
becomes
\be
\label{iHf}
\langle i | H | f\rangle^2  = v_0^2  \left|\sum_n
N^{gcm}_{n_c,n}a_{n,\lambda}\right|^2.
\ee

Having taken care of the definitions of $i$ and $f$ and 
the interaction matrix element, the
remaining task to determine the final state density
$d n_f/ d E$. There are several ways to proceed; we examine two of them.
Method I
is to extend the $f$ channel  basis 
far into the asymptotic region.
Then one can use the $f$-channel eigenfunctions and energies without an
explicit introduction of an RGM wave function.  For a rough estimate,
we can take the energy difference between the eigenstates bracketing
the initial state energy$E_i$,
i.e.
\be
\frac{dn_f}{dE} \approx \frac{1}{(E_\lambda - E_{\lambda-1})}\,\,\,\,{\rm Method~ I}
\ee
where $E_{\lambda -1} < E_i < E_{\lambda}$.

Method II for determining $dn_f/dE$ is to match the
$\psi(z)$ from the GCM to an asymptotic Schr\"odinger wave function 
in the final state.  For the numerical example in Sect II, the asymptotic
wave function is sinusoidal, 
and the match can be carried out with
Eq. (\ref{A}).  The 
resulting density of states is
\be
\frac{dn}{d E} \approx \frac{2 M}{ \hbar^2 \pi k |A(R)|^2}. 
\ee
A big advantage of Method II is that there can be arbitrary
potential interactions in the final state.  The generalization
to arbitrary $V$ is textbook scattering theory.  One first obtains 
the regular and 
irregular wave functions $u(z)$ and $w(z)$ of the scattering equation.
Their relative amplitudes are set so that $w + i u$ is a pure outgoing 
wave.  The GCM wave function $\psi$ is matched to a
linear combination of the two as
\be
\label{fit}
\psi(R) \approx  c_1 u(R) + c_2 w(R).
\ee  Then the density of states is given by
\be
\frac{dn}{d E} \approx \frac{2 M}{ \hbar^2 \pi (c_1^2 + c_2^2)W}
\,\,\,\,{\rm Method~ II}
\ee
where $W = u w' - w u'$ is the Wronskian of the two solutions.

\begin{figure}[tb] 
\begin{center} 
\includegraphics[width=\columnwidth]{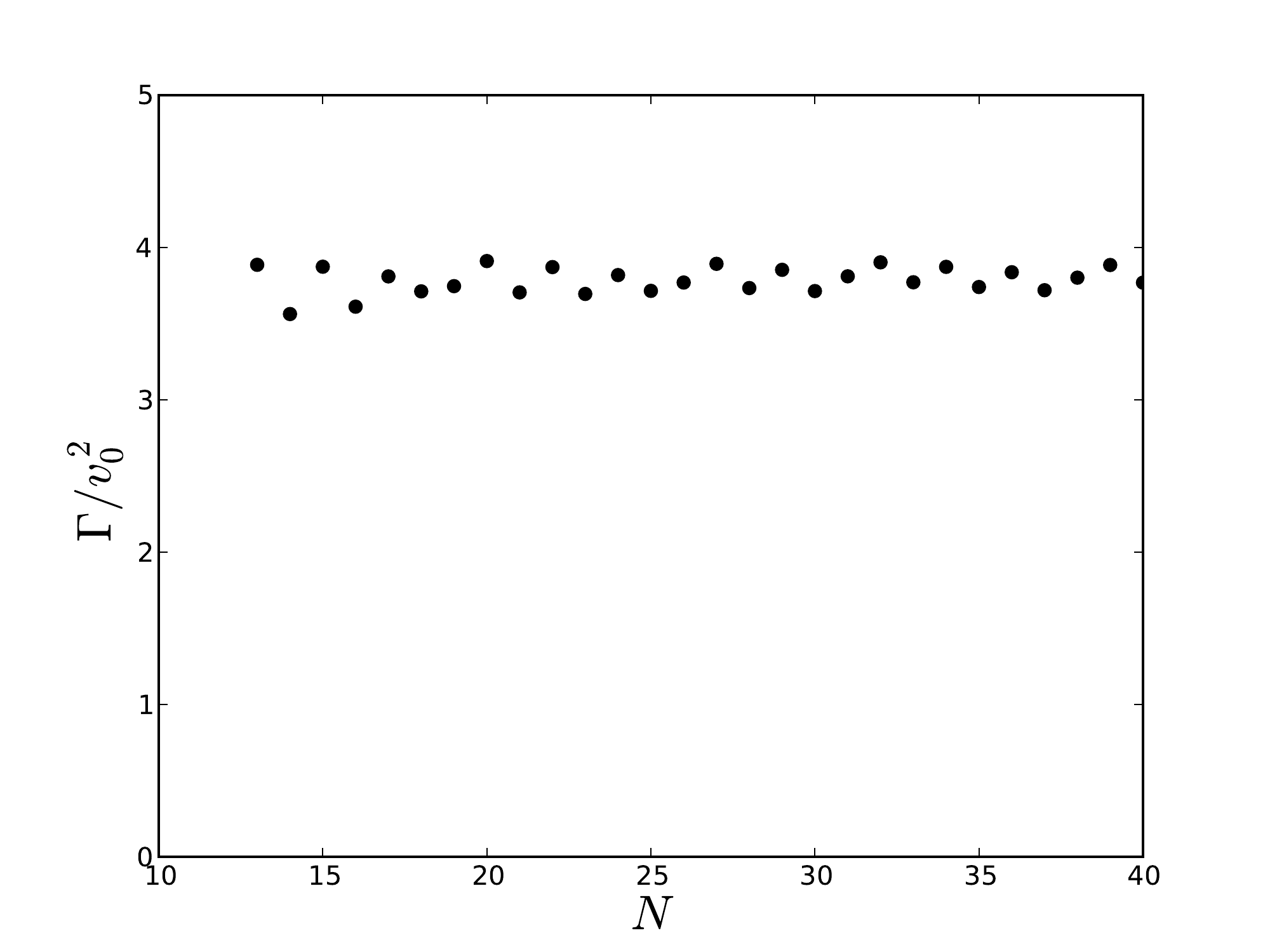}
\caption{$n_c = 4$
\label{g_vs_nd}
Decay width of a configuration $i$ to a channel $f$ in the
potential field Eq. (\ref{rampV}) as a function of the number
of configurations in the $f$ channel.  See text for the
definition of the interaction Hamiltonian.
}
\end{center} 
\end{figure}

We now carry out the numerical solution by the two methods applied to the 
ramp potential Eq (\ref{rampV}).  For this exercise, we
take the interaction
matrix element fom Eq.~(\ref{iHf}) placing the interaction point
in the middle of the ramp, $z = z_i =  -3\Delta z$.  We assume that
the energy of the initial state is $E_i = F z_i$. 
For method I, we start with a
basis of $N=13$ $f$-channel configurations as in the last section.  
More configurations
will be added to the external end of the chain to assess the
convergence of the method.
We don't attempt to tune the GCM Hamiltonian
to produce an eigenstate at $E_i$ but simply
interpolate between the two $f$-channel states
bracketing $E_i$,  i.e. taking weighted average over the two states 
$\lambda,\lambda'$ to estimate $\langle i | H^{gcm} | f \rangle^2$.
The results are shown in Fig. \ref{g_vs_nd}.
One sees that the convergence is quite fast a function of $N$.
For example, the calculated $\Gamma$ with
$N = 13$ configuration is  within 10 \% of the
those calculated at $N=40-41$.
From the systematics, the calculated width can be estimated as
\be
\Gamma_I = (3.8 \pm 0.07) v^2_0.
\ee

For Method II, it is clear from Fig. \ref{rampA} that a 
mesh spacing of $\Delta z = 1$
would not permit a good estimate of the decay width.  
As shown in Fig. \ref{rampA}, reducing 
$\Delta z$ by a factor of 2 permits
a much more accurate estimation of $A$.  Using that mesh
spacing the calculated decay width by Method II is
\be
\Gamma_{II} = (3.77 \pm 0.06) v^2_0.
\ee
We conclude that the two methods agree within their uncertainty and
are accurate to a few percent for the chosen parameters.

\section{Concluding remarks}

It appears to us that GCM is a viable calculational framework 
in reaction theory
involving composite particles as reaction partners.  
With the GCM, one can construct discrete configurations representing
internal excitations of the clusters as well as the 
approximate channel states associated
with decays into smaller clusters.

The most critical approximation is the factorizability in the GCM
of internal and cm wave functions, 
Eq. (\ref{separable}).  This has a direct impact on the kinetic
Hamiltonian. It was found in an early study of the GCM method \cite{pe62} 
it was
found that the calculated 
overall inertial mass of a composite particle may be incorrect.  
The problem doesn't arise in the
present treatment because the factorizability  Ansatz
permits the kinetic operator to be evaluated 
in both the single-particle coordinate representation and in the
representation with the explicit cm coordinate.  
It might not be a good approximation
in practice if there are 
important contributions to the GCM configuration from excited internal states
having different energies and cm wave functions.  However, if the energies
are very different, an even more fundamental assumption is violated.
Namely, it would call into question the utility of the 
mean-field approximation to provide 
a good description of the structure and energy of the lowest  
internal state.  We note that there is also an extensive literature
for dealing with the cm wave function in mean-field theory; see for
example Ref. \cite{gl74}.
Obviously,  more study is needed to determine how reliable the
Ansatz is.

As presented here, a severe limitation of the GCM method is that
the $f$-channel configurations should have energies
that don't vary much from each other on an scale set by the
zero-point cm kinetic energies.  In principle,
this can be ameliorated by including in some way the kinetic energy
into the GCM constraints.
This is can be implemented by
constraining the expectation of the momentum operator 
$p_{rel} = \partial/i \partial z_{rel}$
(as well as $z_{rel}$) in
constructing the configurations.  This requires modifying the
GCM machinery to deal with complex arithmetic, but that should be
a straightforward task.  It has also been suggested to project
on states of good momentum\cite{pe62,wo72}, but the procedure
is challenging from a computational point of view.

There are two distinct regions where the theory of decay widths
might be applied.  At low energies, one might expect that the
internal states are more widely spaced than their decay widths.
In this weak-coupling limit, Eq. (1) can be applied to the
individual resonances.  At higher energies and under certain
conditions on the Hamiltonian, the decay widths may exceed
the level spacings.  Here the
individual decay widths are not of interest but only
their statistically weighted averages. The
relevant physical quantity now becomes the transmission 
coefficient $T$ between fused and separated clusters.  In the
weak coupling limit it can be expressed as
\be
T = 2 \pi \left\langle\frac{\Gamma}{D}\right\rangle
\ee
where the brackets denote averaging and $D$ is the level spacing.  
For large $\Gamma/D$ the
transmission coefficient approaches its unitary limit of $T=1$ and 
there is no need for high accuracy even in the calculation of the average.

\section{Acknowledgments}  We would like to acknowledgment discussions
with T. Kawano and L. Robledo on the physics of nuclear fission, 
motivating some of the questions addressed in this work. 

\section{Appendix: example of an $f$ channel chain}

Extension of the $f$-channel chain into region $A$ requires finding the
configurations that have the largest off-diagonal Hamiltonian matrix elements
to the chain.  This problem was studied in
Ref. \cite{de75} for the nuclear reaction $^{16}$O + $^{16}$O
$\rightarrow$ $^{32}$S.  The
authors assumed axial symmetry in constructing of the basis.
They found that only one particular configuration
in the fused system had a large Hamiltonian matrix element.  The
character of that configuration can be understood in terms of the
orbital fillings with respect to $J_z$, as was carried out in Ref.
\cite{be16}.  We summarize the argument here.  Each oxygen configuration
is constructed from the elementary shell
model, filling the lowest $s$- and $p$-shell orbitals.  The orbitals are
assumed to be independent of nucleon spin and isospin. Thus, each spatial orbital is occupied by 4 nucleons.
The $s$ and the three $p$ orbitals are classified by the angular
momentum $L_z$ about the $z$ axis;
the occupation numbers are $(8,4,4)$ for $L_z =
(0,+1,-1)$.  These occupancies are doubled for two oxygen nuclei
aligned along the $z$ axis.  Thus we seek configurations in the
sulfur nucleus having occupancies $(16,8,8,0,...)$ for $L_z=
(0,+1,-1,+2,-2,...)$.  We have added $L_z = \pm 2$ to the list
because the lowest configurations in that nucleus begin to fill
the $d$ shell.  In addition to the angular momentum quantum number,
the orbitals can be considered to have a good parity.  In the
initial configuration there is the same number of particles in 
each parity orbital.  Thus, 8 of the 16 $L_z=0$ orbitals are
even parity and 8 are odd, and so on.  The combined quantum
numbers are preserved in the $f$ chain, and the possible
configurations in the combined system are quite limited.  Following
the simple shell model, the configuration satisfying these fillings
has completely filled $s$ and $p$ shells, 12 particles in the
$sd$ shell, and 4 particles in the next higher shell.  The
Hamiltonian matrix elements of the external $f$ chain to this
configuration is found to be orders of magnitude larger than
to other configurations of the combined system \cite{de75}.
So for this case at least a clear separation between the $f$ chain
members and the other configurations is possible.

\end{document}